\documentclass[12pt]{iopart}
\usepackage{graphicx}
\begin{document}

\title{Pressure-driven phase transitions in
correlated systems
}

\author{Yu-Zhong Zhang$^1$, Ingo Opahle$^1$, Harald O. Jeschke$^1$, Roser Valent{\'{\i}}$^1$}

\address{$^1$ Institut f\"ur Theoretische Physik, Goethe-Universit\"at  Frankfurt,
Max-von-Laue-Strasse 1, 60438 Frankfurt am Main, Germany}

\ead{valenti@itp.uni-frankfurt.de}

\begin{abstract}
 Motivated by recent experimental measurements on pressure-driven phase transitions
in Mott-insulators as well as the new iron pnictide superconductors,
we show that
 first principles Car-Parrinello molecular dynamics calculations are a powerful
method to describe the microscopic origin of such transitions. We
present results for (i) the pressure-induced insulator to metal
phase transition in the prototypical Mott insulator TiOCl as well as
(ii) the pressure-induced structural and magnetic phase transitions
in the family of correlated metals $A$Fe$_2$As$_2$ ($A$=Ca,Sr,Ba).
Comparison of our predictions with existing experimental results
yields very good agreement.
\end{abstract}

\submitto{\JPCM}

\pacs{71.15.Pd,74.62.Fj,61.50.Ks,74.70.-b}

\noindent(Some figures in this article are in colour only in the
electronic version)

\maketitle

\normalsize
\section{Introduction}

Understanding the microscopic origin of phase transitions in
correlated systems is a challenging task both for theory and
experiment due to the large number of degrees of freedom
(electronic, lattice and orbital) involved in the transitions.  In a
Mott insulator a transition from an insulating to a metallic or even
a superconducting state can be induced by doping the system with
carriers as is the case of the high-$T_c$ cuprates~\cite{Bednorz},
or by applying pressure as in the $\kappa$-charge transfer salts
~\cite{Kurosaki,Kandpal09}. Pressure-driven Mott-insulator to metal
transitions have also been discussed in a few transition metal
oxides like MnO~\cite{Kasinathan0}, V$_2$O$_3$~\cite{Austin}, and
lately in TiOCl~\cite{Kuntscher2,Forthaus}.  Very recently, high
temperature superconductivity has been achieved in a new class of
correlated metals, the iron pnictide compounds, both by doping as
well as by setting the systems under
pressure~\cite{Kamihara,Torikachvili}. The origin of this phase
transition is still under strong debate.

External pressure-induced phase transitions are especially attractive
for microscopic modelling since the chemistry of the system remains
untouched, i.e. possible disorder as in the doping case is almost
inexistent, and one can concentrate on the lattice changes induced by
pressure.  A reliable microscopic description of such transitions
requires a simultaneous treatment of the electronic and lattice
degrees of freedom as is realized in the first principles
Car-Parrinello molecular dynamics method~\cite{CarParrinello}.
Instead of considering the motion of the nuclei and the solution of
the Kohn-Sham equation for the electrons at fixed atomic positions as
separated problems, the Car-Parrinello method explicitly views the
electronic states as dynamical variables and considers these as a
unified problem. By writing a fictitious Lagrangian for the system
which leads to coupled equations of motion for both ions and
electrons, the lattice dynamics and electronic properties can be
treated on the same footing. Furthermore, the Parrinello Rahman
approach to molecular dynamics simulations at constant
pressure~\cite{Rahman} treats the lattice parameters as additional
degrees of freedom, allowing the first principle observation of
structural changes as a function of pressure.  The great advantage of
this procedure is the ability to deal with crystals with anisotropic
compressibilities along different directions as is the case of TiOCl
and the new iron pnictide superconductors. In the present work we will
show by reviewing recent calculations on pressure-driven transitions
in the multiorbital Mott insulator TiOCl as well as the family of iron
pnictides $A$Fe$_2$As$_2$ ($A$=Ca,Sr,Ba) that the Car-Parrinello
method is often superior to the more common Born-Oppenheimer methods
for describing the features of the phase transitions.

\section{Overview on $\mathrm{TiOCl}$ and $A\mathrm{Fe}_2\mathrm{As}_2$}

\begin{figure}[tb]
\centering{\includegraphics[width=0.50\textwidth]{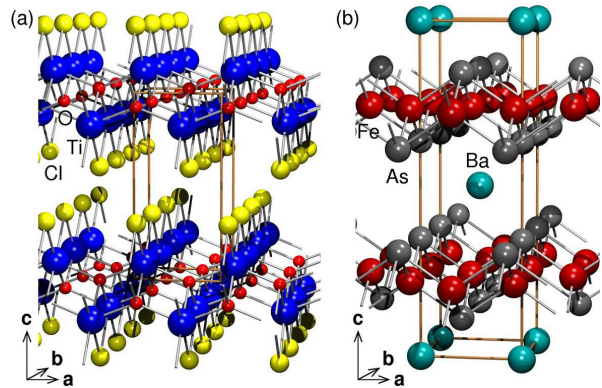}}
\caption{Lattice structures of (a) TiOCl and (b) $A$Fe$_2$As$_2$
($A$=Ca,Sr,Ba) for the example of BaFe$_2$As$_2$.}
\label{fig:struct}
\end{figure}

TiOCl and AFe$_2$As$_2$ (A=Ca,Sr,Ba) are both layered compounds
containing $3d$ transition metal ions as shown in
Fig.~\ref{fig:struct}.  Correlation among the open $d$ shell
electrons is responsible for a rather unconventional behavior in
these
materials~\cite{Kuntscher2,Forthaus,Torikachvili,Seidel,Hoinkis,Shaz,Krimmel,Rueckamp,Abel,Sing,
  ZKJSV,ZJV2,
  Blanco-Canosa,Kuntscher1,Huang,Rotter1,Goldman,Zhao,Yu,Rotter2,
  Sasmal,Ahilan,Saha,Kreyssig,Goldman2,Kumar,Lee,Alireza,
  Colombier,Kimber,Fink}.  In TiOCl, titanium (Ti) and oxygen (O) form
bilayers stacked along the $c$ direction and separated by chlorine (Cl)
layers, while in $A$Fe$_2$As$_2$ ($A$=Ca,Sr,Ba), layers of Fe-As
tetrahedra are separated by alkaline earth metal ions.

\subsection{TiOCl}

TiOCl is a quasi-one-dimensional Mott insulator at room temperature
where Ti chains along the $b$ direction are weakly
coupled~\cite{Seidel,Hoinkis}. Upon lowering the temperature, TiOCl
undergoes two consecutive phase transitions from a Mott insulating
state to a spin-Peierls phase with dimerized Ti chains through an
intermediate structural incommensurate phase with dominant
out-of-chain incommensurability  even under zero external magnetic
field~\cite{Shaz,Krimmel,Rueckamp,Abel}. Since the discovery of this
unusual intermediate phase, many efforts have been devoted to
understanding its microscopic origin
~\cite{ZJV1,Mastrogiuseppe2,Mastrogiuseppe3,Pisani}. Recently, a
two-dimensional frustrated spin-Peierls model was proposed, where
the model parameters were calculated from \textit{ab initio} density
functional theory (DFT)~\cite{ZJV1}. Besides the dominating
interactions along chains, it was pointed out that the interchain
magnetic frustrations may play an important role in the formation of
the unconventional incommensurate phase in addition to the
interchain elastic couplings.

Due to the $3d^1$ configuration of Ti$^{3+}$ and quenching of the
orbital degrees of freedom~\cite{Rueckamp,dasgupta}, TiOCl was
thought to be a model system for the one band Hubbard model.  In
order to investigate a possible Mott insulator-to-metal transition
in this system in the framework of the one-band Hubbard model
various experimental groups have recently investigated the effect of
pressure ~\cite{Kuntscher2,Forthaus,ZJV2,Blanco-Canosa,Kuntscher1}
and electron doping~\cite{Sing,ZKJSV} on TiOCl.  Interestingly,
pressure driven experiments lead to different conclusions.
Kuntscher {\it et
  al.}~\cite{Kuntscher2,Kuntscher1} reported from optical measurements
and X-ray diffraction analysis the observation of an
insulator-to-metal transition at 16~GPa accompanied by a structural
symmetry change.  Forthaus {\it et al.}~\cite{Forthaus}, on the other
hand, neither observed any indication of a metallic state up to
pressures of 24~GPa from transport measurements nor did they identify
from X-ray diffraction analysis changes on the crystal symmetry at
16~GPa. Recently, Blanco-Canosa {\it et al.}~\cite{Blanco-Canosa}
found a dimerized insulating state at high pressure by X-ray
diffraction analysis and magnetization measurements combined with {\it
  ab initio} DFT results. The DFT calculations in Ref.~\cite{Blanco-Canosa}
consisted in performing a structural optimization within the
generalized gradient approximation (GGA) and an analyis of the
electronic properties within GGA+U.  It should be stressed that GGA
gives a metallic state for TiOCl while GGA+U describes TiOCl as an
insulator.  In the following, we will argue that inclusion of strong
correlation is crucial for obtaining the correct optimized lattice
structure.  Therefore, the theoretical understanding of the
dimerization under pressure~\cite{Blanco-Canosa} may be
 questionable due to the alternate
use of different functionals in one study.

In a recent work~\cite{ZJV2}, we resolved the previous inconsistencies
by performing {\it ab initio} Car-Parrinello molecular
dynamics~\cite{CarParrinello,Rahman} within the projector-augmented
wave basis (CP-PAW)~\cite{Bloechl} on TiOCl under pressure.  The GGA+U
functional was considered throughout the whole study.  Within GGA+U
correlation effects are included at the mean-field level. The
simulations show that two phase transitions occur; a first one from
the Mott insulating state to an intermediate dimerized metallic state
accompanied by a symmetry lowering from orthorhombic $Pmmn$ to
monoclinic $P2_1/m$, and a second transition to a normal metal with
recovered $Pmmn$ symmetry. The first phase transition is consistent
with the observations of Kuntscher {\it et
  al.}~\cite{Kuntscher2,Kuntscher1} and the structural phase
transition is in agreement with the experimental result of
Ref.~\cite{Blanco-Canosa}. The second phase transition has not yet
been reported experimentally. Moreover, the pressure effects involve
the three $t_{2g}$ orbitals in TiOCl.

\subsection{ $A\mathrm{Fe}_2\mathrm{As}_2$}

The materials $A$Fe$_2$As$_2$ ($A$=Ca,Sr,Ba) are paramagnetic metals
at room temperature. With decreasing temperature, a magnetic phase
transition to a stripe-type antiferromagnetic (AF) ordered state and
a structural phase transition from tetragonal to orthorhombic
symmetry take place
simultaneously~\cite{Huang,Rotter1,Goldman,Zhao}.  Even more
exciting is the fact that these systems are superconductors both
under
pressure~\cite{Torikachvili,Kumar,Lee,Alireza,Colombier,Kimber} and
electron or hole doping~\cite{Rotter2,Sasmal,Ahilan,Saha}, which has
attracted considerable and persisting interest.

Since doping usually induces disorder, it is more practical to
investigate the origin of the phase transition and the phenomenon of
superconductivity by applying external pressure. However, while the
pressure-driven magnetic and structural phase transitions in
CaFe$_2$As$_2$ have been thoroughly
investigated~\cite{Kreyssig,Goldman2}, those in SrFe$_2$As$_2$ and
BaFe$_2$As$_2$ have not yet been fully studied due to the high Sn
content that the grown crystals have~\cite{Kumar}. Therefore, a
theoretical prediction of the phase transitions in SrFe$_2$As$_2$
and BaFe$_2$As$_2$ is desirable since it is free of the Sn-problem.
Unfortunately, it is well-known that the magnetization and the
volume of the iron pnictide compounds are overestimated by DFT
calculations within GGA when structural optimizations are
performed~\cite{Mazin,ZKOJV,Opahle09}. We will show here that due to
this overestimation we only have to simulate higher pressures than
the experimental ones in order to reach the experimental volume
conditions while structural changes at critical pressures and the
main features of the electron properties within different phases can
still be captured properly.

Optimization of cell parameters and atomic positions under pressure
within the framework of density functional theory (DFT) has been done
by Yildirim~\cite{Yildirim} and Xie {\it et al.}~\cite{Xie}.  By
considering Vanderbilt-type ultrasoft pseudopotentials, a smooth
structural transition for CaFe$_2$As$_2$ was observed under pressure
without detection of a sudden sizable increase of the cell parameters
$a$ and $b$ or a strong decrease of the cell parameter $c$ which is
inconsistent with experimental results~\cite{Kreyssig}. Xie {\it et
  al.}~\cite{Xie} optimized within the full potential linearized
augmented plane wave method (FPLAPW) the orthorhombic lattice
structure for BaFe$_2$As$_2$ under pressure by relaxing the internal
parameter $z_{As}$ and the $c/a$ ratio while keeping the $b/a$ ratio
fixed. This procedure doesn't allow for the detection of the
structural and magnetic phase transitions.

We investigated the phase transitions of $A$Fe$_2$As$_2$
($A$=Ca,Sr,Ba) under pressure in the lower temperature region with the
CP-PAW method within the spin-polarized GGA functional.  The first
order phase transition from an orthorhombic phase to a collapsed
tetragonal phase in CaFe$_2$As$_2$ is confirmed with relative changes
of lattice parameters, bond lengths and angles agreeing very well with
the experimental observations, indicating the validity of the
application of the CP-PAW method to this system. At the critical
pressure, an abrupt disappearance of magnetization is also found as
observed experimentally. In particular, our calculations can account
for the sudden expansion in the $ab$ plane observed
experimentally~\cite{Kreyssig} at elevated pressure crossing the
critical pressure as we discussed in ref.~\cite{ZKOJV}, which was not
obtained in other calculations. Applying this method to study
SrFe$_2$As$_2$ and BaFe$_2$As$_2$, where less is known about the
features of the phase transition, a simultaneous structural
(orthorhombic to tetragonal) and magnetic phase transition is
predicted at high pressure. However, we observe a weak first-order
phase transition for SrFe$_2$As$_2$ and a continuous phase transition
for BaFe$_2$As$_2$ in contrast to the strongly first order phase
transition in CaFe$_2$As$_2$.  Finally, it should be pointed out that
multi-orbital physics is important in $A$Fe$_2$As$_2$ ($A$=Ca,Sr,Ba)
since all five $d$ orbitals cross the Fermi level.

\section{Details of our calculations}

In our studies of both systems, full, unbiased relaxations of all
lattice and electronic degrees of freedom have been performed at each
pressure value.  In TiOCl, local coordinates are chosen as $x=b$, $y=c$,
$z=a$ while in $A$Fe$_2$As$_2$ ($A$=Ca,Sr,Ba), $x=a$, $y=b$, $z=c$.

\begin{figure}[tb]
\centering{\includegraphics[width=0.9\textwidth]{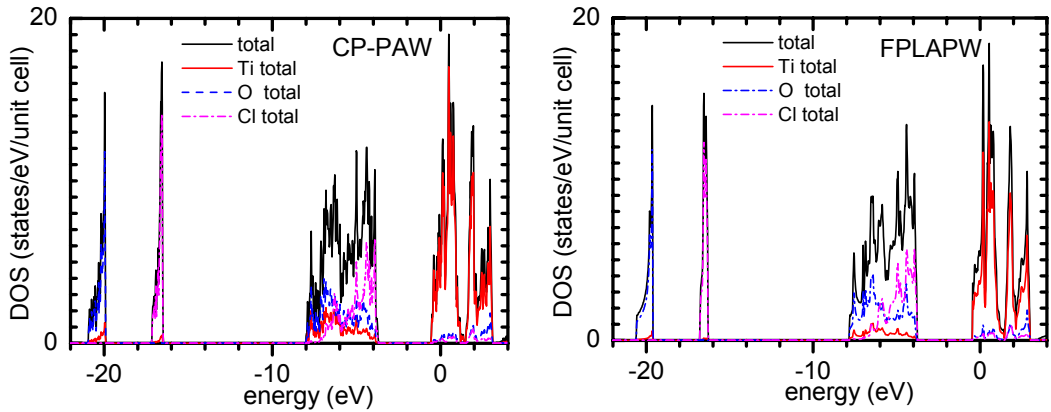}}
\caption{Comparison of the density of states for TiOCl between
CP-PAW (a) and FPLAPW (b) methods within GGA.}
\label{fig:ComparisonPAWbasis}
\end{figure}

All our calculations were performed with time steps of 0.12~fs at
zero temperature. The system size was $4\times 4 \times 4$ $k$
points for $A$Fe$_2$As$_2$ ($A$=Ca,Sr,Ba) and $6\times 6 \times 6$
for TiOCl on doubled unit cells corresponding to different AF
configurations~\cite{ZJV2,ZKOJV}. We used high energy cutoffs of
612~eV and 2448~eV for the wave functions and charge density
expansion, respectively. The total energies were converged to less
than 0.01~meV/atom and the cell parameters to less than 0.0005~\AA.
In TiOCl, a value of $U=1.65$~eV was used for the GGA+U calculations
with which correct spin exchange along $b$ is
reproduced~\cite{ZJV1}. In $A$Fe$_2$As$_2$ ($A$=Ca,Sr,Ba), the
3s3p3d (4s4p4d/5s5p5d) states in Ca (Sr/Ba) are treated as valence
states and 3d4s4p in Fe and As, while in TiOCl, the 3s3p3d4s in Ti,
2s2p in O and 3s3p in Cl. We have checked our calculations within
GGA based on these choices of valence states with the FPLAPW method
as implemented in the WIEN2k code~\cite{Blaha}. Very good agreement
was found between these two methods, indicating the correctness of
our constructed PAW basis for both systems.  In
Fig.~\ref{fig:ComparisonPAWbasis}, we show one of the comparisons
for TiOCl between CP-PAW and FPLAPW within GGA. It is found that the
total density of states (DOS) and partial DOS (pDOS) calculated from
FPLAPW can be reproduced by CP-PAW. Not only the peak position but
also the shape of the DOS are almost the same throughout the whole
energy range. More spikes in the DOS from CP-PAW come from less $k$
points used in the calculation.

\begin{table}[!ht]
  \caption{Comparison of equilibrium lattice parameters for TiOCl optimized by
    different functionals. B3LYP data were obtained from Ref.~\cite {Pisani}}
\label{abc}
\centering{\begin{tabular}{cccc}
    \hline\hline & a (\AA) & b (\AA) & c (\AA) \\ \hline
    exp. & 3.79 & 3.38 & 8.03 \\
    CRYSTAL B3LYP & 3.81 & 3.49 & 8.69 \\
    GGA & 3.89 & 3.30 & 8.06 \\
    GGA+U(1.65,FM) & 3.80 & 3.49 & 8.12 \\
    GGA+U(1.65,AF) & 3.81 & 3.42 & 7.51 \\ \hline
\end{tabular}}
\end{table}

\begin{table}[!ht]
  \caption{Comparison of equilibrium atomic distances and angles
    for TiOCl optimized by different functionals. B3LYP data obtain from
    Ref.~\cite {Pisani}} \label{dist}
  \centering{\begin{tabular}{ccccc}
      \hline\hline & Ti-O$|_{a}$ (\AA) & Ti-Cl (\AA) & Ti-O-Ti$|_{a}$ &
      \\ \hline
      exp. & 1.96 & 2.40 & 150$^\circ$ &  \\
      CRYSTAL B3LYP & 1.98 & 2.45 & 148.6$^\circ$ &  \\
      GGA & 1.98 & 2.48 & 156.4$^\circ$ &  \\
      GGA+U(1.65,FM) & 1.97 & 2.44 & 149.1$^\circ$ &  \\
      GGA+U(1.65,AF) & 1.97 & 2.41 & 151.4$^\circ$ &  \\ \hline
\end{tabular}}
\end{table}

Now let us investigate the effect of strong correlations on the
lattice optimization\cite{Pisani_opt} of TiOCl. In Table.~\ref{abc}, we present the
optimized lattice parameters within different functionals compared to
the experimental equilibrium state in TiOCl. For the lattice parameter
$c$, we find that it is rather hard to determine precisely due to the
fact that there only exist weak van der Waals forces between layers
along $c$. Thus, by varying the lattice parameter $c$, the total
energy remains almost unchanged. For the lattice parameters $a$ and
$b$, B3LYP and GGA+U with both FM and AF configurations can provide a
better comparison to experiment than GGA does, indicating the
importance of strong correlations which are absent in GGA. Futhermore,
it is found that the results from GGA+U with an underlying AF ordering
is even better than that from GGA+U with FM ordering, which indicates
that the correct setting of the spin configuration improves the
lattice optimization. In Table.~\ref{dist}, we show the optimized
atomic distances and angles by different functionals compared to the
experimental results. Again, the same trend is found as in
Table.~\ref{abc}, {\it i.e.}, with the inclusion of correlations as
implemented in GGA+U and in B3LYP, the optimized structures are closer
to the experimental one.

In contrast to TiOCl, the iron pnictide compounds are metallic even
in the AF state and are considered as moderately correlated systems.
Therefore, the lattice optimizations for $A$Fe$_2$As$_2$
(A=Ca,Sr,Ba) have been performed within GGA.

\section{Results}

\subsection{TiOCl under pressure}

In the study of TiOCl under pressure, we use a renormalized pressure
due to the overestimation of the critical value. However, it has a
different origin compared to that in $A$Fe$_2$As$_2$ (A=Ca,Sr,Ba). As
reported from experiments~\cite{Hoinkis}, TiOCl is a Mott insulator at
room temperature and ambient pressure. Only short-range AF spin
fluctuations exist in TiOCl. However, to reach the insulating state by
DFT calculations, one has to impose certain long-range spin ordering.
It is well-known~\cite{Kotliar} that the AF insulator is more robust
than the paramagnetic Mott insulator under bandwidth control; one has
to apply much higher pressure to induce an insulator-to-metal
transition if long-range AF ordering is present.

\begin{figure}[tb]
\centering{\includegraphics[width=0.9\textwidth]{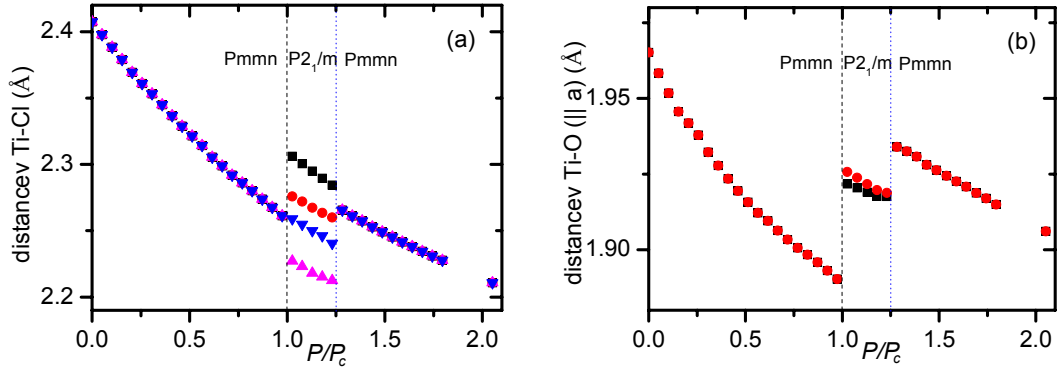}}
\caption{Calculated change of atomic distances of TiOCl as a
function
  of applied external pressure, normalized to the critical pressure.
  (a) The distance between Ti and Cl. (b) The distance between Ti and
  O along $a$.} \label{fig:disPTiOCl}
\end{figure}

In Fig.~\ref{fig:disPTiOCl} we present the calculated change of atomic
distances as a function of applied external pressure. The lattice
parameters at each pressure are obtained by fully relaxing the lattice
structure without any symmetry constraints. Although a magnetic
ordering has to be imposed within the GGA+U functional (AF ordering
along $b$ and FM configuration along the other directions) the results
are completely consistent with the
experiments~\cite{Forthaus,Kuntscher1}.

At $P=P_{c}$ (black dashed vertical line in
Fig.~\ref{fig:disPTiOCl}), the equivalent Ti-Cl distances in the
lower pressure region split into four different values (see left
panel) and the Ti-O distance into two (see right panel), indicating
a lowering of the symmetry. We found that the space group changes
from $Pmmn$ to $P2_1/m$ which agrees with the experimental
observation~\cite{Blanco-Canosa}. Most interestingly, both
inequivalent Ti-O distances show jumps at this phase transition,
indicating that the system expands along $a$ in agreement with
experiment~\cite{Kuntscher3}. Further increasing the external
pressure, we find a second jump of the atomic distances at $P=P'_c >
P_c$ (blue dotted vertical line in Fig.~\ref{fig:disPTiOCl}). At
this pressure, all the inequivalent Ti-Cl and Ti-O distances become
equivalent again, indicating the recovery of the $Pmmn$ symmetry.
This phase transition has not yet been observed.

\begin{figure}[tb]
\centering{\includegraphics[width=0.9\textwidth]{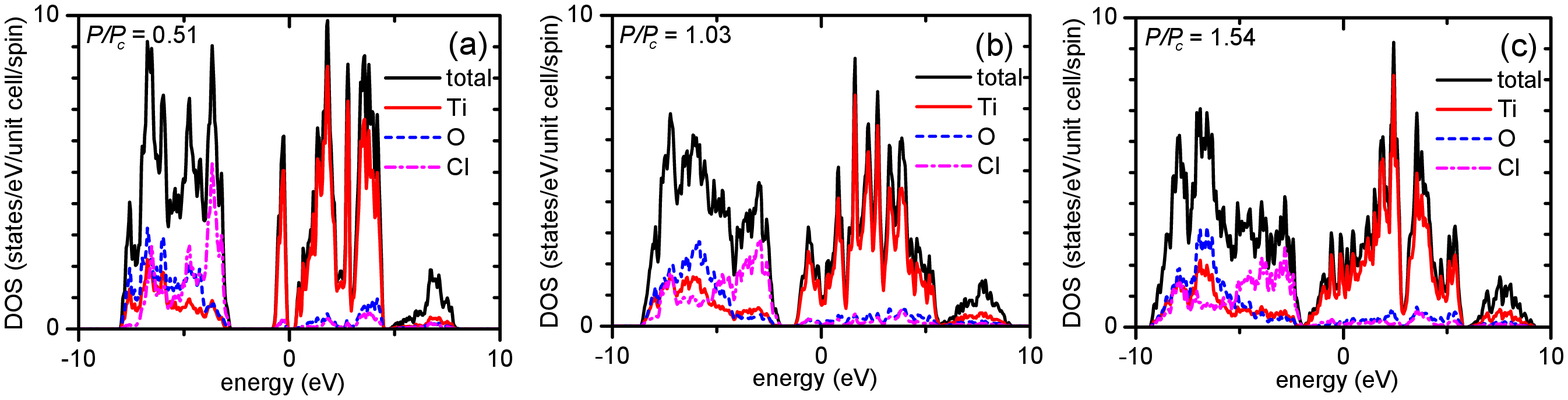}}
\caption{Density of states of TiOCl in three different phases. (a)
is
  for the low pressure Mott insulating state ($P < P_c$), (b) for the
  dimerized metallic phase ($P_c < P < P'_c$), (c) for the uniform
  metal ($P > P'_c$).}
\label{fig:DOSresTiOCl}
\end{figure}

Fig.~\ref{fig:DOSresTiOCl}) presents the DOS in these three phases.
Fig.~\ref{fig:DOSresTiOCl})~(a) shows the DOS for the lower pressure
case. The DOS looks very similar to that at ambient pressure except
for the widening of the bands and the reduction of the gap
amplitude. Right above the first phase transition, the gap is closed
as shown in Fig.~\ref{fig:DOSresTiOCl})~(b), which means that the
system should show metallicity. The obtained insulator-to-metal
transition is consistent with optical
experiments~\cite{Kuntscher2,Kuntscher1} but contradicts the
electrical resistivity measurements~\cite{Forthaus}. A possible
reason is that TiOCl is a strongly anisotropic crystal with a large
van der Waals gap along $c$. Thus, if the resistivity measurements
have been done along $c$, the system should show insulating
behavior, reminiscent of cuprates. Further increasing the pressure
to $P > P'_c $, it is shown in Fig.~\ref{fig:DOSresTiOCl})~(c) that
the metallicity remains with further widening of the bands and even
nearly closure of the gap between the $p$ states of O, Cl and the
$d$ states of Ti. Therefore, this phase transition is exclusively
induced by the structural change.

\subsection{$A$Fe$_2$As$_2$ under pressure}

\begin{figure}[tb]
\centering{\includegraphics[width=0.9\textwidth]{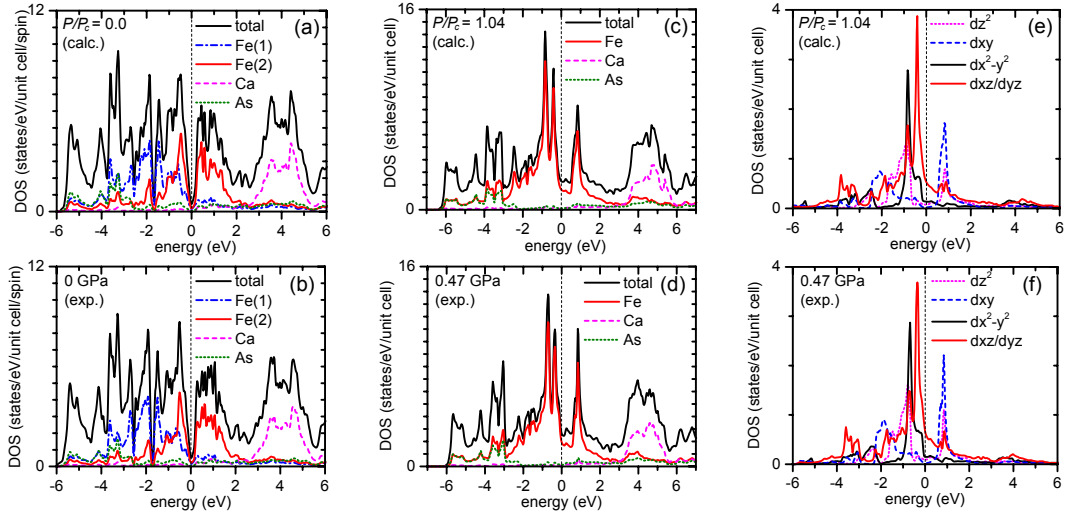}}
\caption{Comparison of DOS for CaFe$_2$As$_2$ calculated for
optimized
  and experimental lattice structures in both AF orthorhombic phase
  ((a), (b)) and paramagnetic 'collapsed' tetragonal phase ((c), (d))
  at low temperature. More detailed comparison of the pDOS of Fe $d$
  states is shown in (e) and (f).}
\label{fig:ComDOS122v}
\end{figure}

In Fig.~\ref{fig:ComDOS122v}, we show the comparison of densities of
states calculated based on optimized and experimental lattice
structures for CaFe$_2$As$_2$ in both the AF orthorhombic phase and
the paramagnetic 'collapsed' tetragonal phase at low temperature.
Figs.~\ref{fig:ComDOS122v}~(a) and (b) show comparison of DOS in the
orthorhombic phase. It is found that the DOS are almost the same
although the volume and the magnetic moment are both overestimated by
DFT calculations within the spin-polarized GGA functional.
Figs.~\ref{fig:ComDOS122v}~(c), (d) are the comparison of DOS in the
'collapsed' tetragonal phase, and Figs.~\ref{fig:ComDOS122v}~(e), (f)
the corresponding comparison of pDOS of Fe $d$ states. It is obvious
that both DOS and pDOS are almost identical. Peaks coming from Fe
$d_{xz}/d_{yz}$ and $d_{x^2-y^2}$ are shifted away from the Fermi level
due to the structural change under pressure which avoids the high
instability at the Fermi level present in the high-temperature
tetragonal phase. Considering the fact that the change of the lattice
structure under elevated pressure observed experimentally can be
described well by our CP-PAW calculations, in particular the jumps of
the volume and the distances of Fe-As at the phase transition which are the
two key quantities determining the volume collapse and the electronic
properties, respectively, (as shown in Ref.~\cite{ZKOJV}), we come to
the conclusion that our DFT structural optimizations are valid in both
low temperature phases, namely, the AF orthorhombic phase and the
'collapsed' tetragonal phases, except that we have to apply higher
pressure than the critical value obtained from experiments to
eliminate the overestimated magnetic moment and volume at ambient
pressure and to reach the experimental volume conditions. In order to
avoid confusion, we renormalized the pressure with respect to the
critical value throughout this work.

\begin{figure}[tb]
\centering{\includegraphics[width=0.9\textwidth]{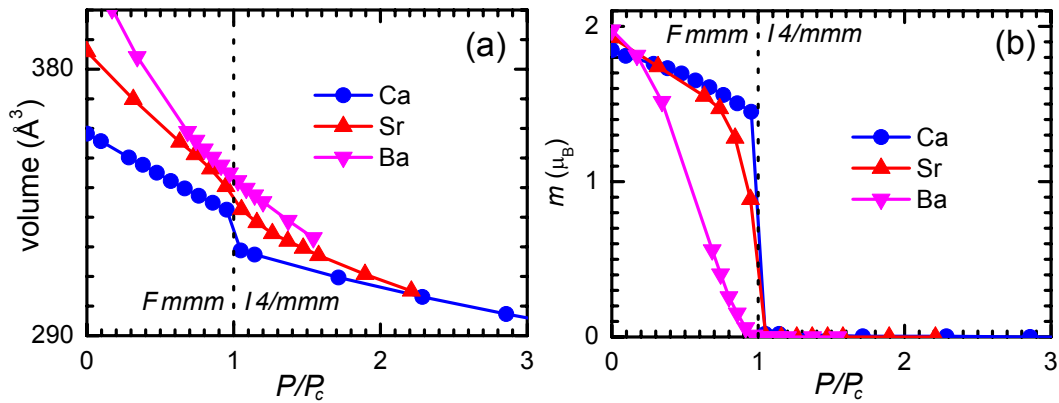}}
\caption{Calculated changes of (a) volume and (b) magnetization of
  $A$Fe$_2$As$_2$ ($A$=Ca,Sr,Ba) as a function of pressure, normalized
  to the corresponding critical pressures.}
\label{fig:StructMag122}
\end{figure}

In Figs.~\ref{fig:StructMag122}~(a) and (b) we present the
calculated changes of the volume and magnetization of
$A$Fe$_2$As$_2$ ($A$=Ca,Sr,Ba) as a function of pressure. In the
case of CaFe$_2$As$_2$, the volume and the magnetization decrease
gradually with increasing pressure and show a discontinuity at the
critical pressure, where the system undergoes simultaneously the
structural and magnetic phase transitions from an AF orthorhombic
phase to a volume 'collapsed' paramagnetic phase.  Our results are
in very good agreement with experimental data~\cite{Kreyssig} with a
volume collapse of $\Delta V^{\rm th} \approx 4.1\%$, $\Delta V^{\rm
exp} \approx 4\%$.

In Fig.~\ref{fig:StructMag122}, we also show the predictions for
SrFe$_2$As$_2$ and BaFe$_2$As$_2$.  We find that the structure and
the magnetic phase transitions under pressure still occur
simultaneously in these two compounds, but the nature of the phases
is distinctly different from the CaFe$_2$As$_2$ case. This has not
yet been confirmed experimentally due to the problem of high Sn
content~\cite{Kumar} in the samples. SrFe$_2$As$_2$ shows smaller
magnetization and volume jumps at the critical pressure compared to
CaFe$_2$As$_2$ while we can hardly detect any discontinuity in
BaFe$_2$As$_2$, which in this respect resembles more
closely~\cite{Opahle09} the parent compound of the iron pnictide
superconductors, LaFeAsO.

In summary, by considering {\it ab initio} Car Parrinello molecular dynamics
calculations, we were able to describe the microscopic details of the pressure-induced
phase transitions in TiOCl and the family $A$Fe$_2$As$_2$ ($A$=Ca,Sr,Ba)
with very good agreement to experimental observations.  The
CP-PAW proves to be a reliable and powerful
method to describe structural and magnetic changes
at the phase transition.

\subsection{Acknowledgments}

We thank the Deutsche Forschungsgemeinschaft for financial support
through the TRR/SFB 49 and Emmy Noether programs.

\end{document}